\documentclass[letterpaper]{scrartcl}
\usepackage{fullpage}
\usepackage{listings}
\usepackage{alltt}
\usepackage{multirow}
\usepackage{epsfig}
\usepackage[hyphens]{url}
\usepackage[top=1in, bottom=1.2in, left=1in, right=1in]{geometry}

\usepackage{booktabs} 
\usepackage{listings}
\usepackage{xcolor}
\usepackage{multicol}
\usepackage{graphicx}
\usepackage{amssymb}
\usepackage{pifont}
\newcommand{\xmark}{\ding{55}}
\newcommand{\cmark}{\ding{51}}

\renewcommand{\paragraph}[1]{\vspace{.2em} \noindent\textbf{#1}}

\lstdefinestyle{c}{
  language=java,
  basicstyle=\ttfamily,
  keywordstyle=\color{magenta},
  stringstyle=\color{blue},
  commentstyle=\color{black!50}
}

\newfont{\secfntab}{ptmb8t at 10pt}

\begin{document}
\title{Hey, you, keep away from my device: remotely implanting a virus expeller to defeat Mirai on IoT devices}

\author{
Chen Cao$^{\dag\ddag}$, Le Guan$^{\ddag}$,
Peng Liu$^{\ddag}$,
Neng Gao$^{\dag}$,
Jingqiang Lin$^{\dag}$,
Ji Xiang$^{\dag}$ \\
\textit{Institute of Information Engineering, CAS, China$^{\dag}$}\\
\textit{College of Information Science and Technology, Penn State University$^{\ddag}$}\\
\texttt{\{caochen11\}@mails.ucas.ac.cn, \{lug14, pliu\}@ist.psu.edu} \\
\texttt{\{gaoneng, linjingqiang, xiangji\}@iie.ac.cn}
\\
\\
}








\date{
June, 2017\\
PSU-S2-TR-2017-04001
}
\maketitle

\begin{abstract}
    \vspace{5pt}
    \begin{center}
        {\Large \textbf{Abstract}}
    \end{center}

    \vspace{5pt}
Mirai is botnet which targets out-of-date Internet-of-Things (IoT) devices. The disruptive
Distributed Denial of Service (DDoS) attack last year has hit major Internet
companies, causing intermittent service for millions of Internet users. Since
the affected devices typically do not support firmware update, it becomes
challenging to expel these vulnerable devices in the wild.

Both industry and academia have  made great efforts in amending the situation.
However, none of these efforts is simple to deploy, and at the same time
effective in solving the problem. In this work, we design a collaborative
defense strategy to tackle Mirai. Our key idea is to take advantage of human
involvement in the least aggressive way. In particular, at a negotiated time
slot, a customer is required to reboot the compromised device, then a
``white'' Mirai operated by the manufacture breaks into the clean-state IoT
devices immediately. The ``white'' Mirai expels other malicious Mirai
variants, blocks vulnerable ports, and keeps a heart-beat connection with the
server operated by the manufacturer. Once the heart-beat is lost, the server
re-implants the ``white'' Mirai instantly. We have implemented a full
prototype of the designed system, and the results show that our system can
evade Mirai attacks effectively.
\end{abstract}

\newpage

\section{Introduction}


In October 21st, 2016, Dyn, an infrastructure vendor, which serves Internet's
top giants, such as Netflix and Twitter, was attacked by the record
Distributed Denial of Service (DDoS) attack~\cite{mirai-ddos}. The attack was later
found originated from a botnet malware -- Mirai. It was the same botnet
malware that attacked an security researcher's blog website and had the record
620 Gpbs stream in September 21st~\cite{mirai-first}. The Mirai malware mainly targets
digital video recorders (DVRs) and IP cameras, which are mainly old and
low-end products which have no firmware update capability. In addition, since
the firmware is read only, the Mirai code can only stay in the DRAM of the
device; rebooting the device also wipes the Mirai code. That being said, once
infected, the only recourse is to wait for these devices to be rebooted since
there is no path to remediation through any type of reconfiguration. Given the
large amount of these vulnerable devices, and the fact that there is no way
patch them, Mirai based attacks have become a time bomb which no one can
defuse。

Under the pressure from media, some manufacturers claimed they were to recall
vulnerable products linked to this massive DDoS attack. For example, Hangzhou
Xiongmai Technology planned to recall 4.3 million Internet-connected camera
products from the U.S market~\cite{mirai-recall}. Although the company invests a huge time and
energy to amend the situation, it only mitigates later attacks, simply because
device users have no incentive to cooperate in the  recall process. They are
not willing to spend  time on packing these devices and sending them back,
because the Mirai malware does not affect the normal operations of the
compromised devices. As a result, there are still a lot of vulnerable devices
remaining in the wild.

When passive recalling turned out to be inefficient, the manufacturer could
also actively contact customers to perform remote diagnose. If the device is
found compromised, the customer  agent could explain the consequence  of
leaving the device in the wild, and urge the customer to replace the device
for free. However, a manufacturer may have millions of products sold around
the world~\cite{mirai-recall}. It is not practical for the customer service  to contact
each user to fix the problem.


Academia is also active in solving the Mirai problem. In~\cite{anti-mirai}, the authors
propose a fancy idea to deploy a ``white'' Mirai-like system. The government
is required to record vulnerable products and push the related manufacturers to
fix the problem. This solution inherits  a bunch of code from Mirai. Notably,
like the malicious Mirai, the ``white'' Mirai actively scans neighbour
vulnerable devices, and infects them. The infected devices then become immune
to any other similar attack, as a result of blocked ports. As the ``white''
Mirai spread itself through infection, it exhausts resources and imposes heavy
overhead for the resource-constraint devices. Furthermore, there is a combat
between the ``white'' and real Mirai. Only the winner takes control of the
device. In fact, Mirai has already infected millions of devices which has a
good chance to win the game. In this sense, the solution is non-deterministic.
Last but not least, the solution has legal concerns. Although this Mirai-like
system is conducted by the government, it is still illegal to compromise a
device without the consent of the user or the manufacturer.

In summary, both industry and academia have no solution that is simple to
deploy, and at the same time effective in addressing the problem.


\begin{table}[t!]
\caption{Comparison of different solutions}
\label{tab:comparison}
\begin{center}
\begin{tabular}{l|c c c }
\hline
           & ~~Simplicity~~ & ~~Effectiveness~~ & ~~Manageableness~~ \\ \hline
~~Recall  & \xmark & \cmark & \cmark \\ \hline
~~Customer Service~~ & \xmark & \cmark & \xmark \\ \hline
~~White Mirai & \cmark & \xmark & \xmark \\ \hline
~~Our solution & \cmark & \cmark & \cmark \\ \hline
\end{tabular}
\end{center}
\end{table}

In this paper, we propose a solution which defeats Mirai effectively and
avoids aforementioned issues. The key idea of our solution is to take
advantage of human involvement in the least aggressive way. The idea of
kicking in human is based on the  observation that rebooting the device also
wipes the Mirai code in memory. Based on this observation, we propose a
collaborative defense strategy to tackle Internet-of-Things (IoT) device based DDoS attacks. At a
high-level, our solution resembles the ``white'' Mirai work in that both
utilize attacker's method as a defense measure for “tit for tat”. However, we
require the customer to collaborate with the manufacture by rebooting the
device at the negotiated time slot. At this time slot, the malicious Mirai has
very little chance to kick in, while the ``white'' Mirai could ``break'' into
the device in time. In particular, the manufacturer builds an implanter server
which remotely implants a virus expeller into a vulnerable device to expel
other Mirai away from it.  Once implanted into a device, the virus expeller
closes all the ports used by malicious Mirai, leaving no room for Mirai to
access this specific device. The implanter server operated by the manufacturer
is responsible for finding the vulnerable devices, implanting the virus
expeller and keeping the virus expeller alive in a specific device.

Table~\ref{tab:comparison} compares our solution with three aforementioned
solutions. In the table, simplicity means there is no much burden for the
customers and the manufacture to carry on the emendation. Effectiveness means
the solution can effectively defeat Mirai. Manageableness means the
manufacturer can keep track of the deployment of the system.

In summary, we make three main contributions in this paper:

\begin{enumerate}
\item We analyze the architecture of a legacy digital video recorder which is subject to the Mirai attack. 


\item We propose a practical solution to defeat Mirai the botnet malware, which is simple, effective and manageable.

\item We implement a proof-of-concept prototype and show that it can successfully shield vulnerable devices from Mirai's infection. 
\end{enumerate}

\paragraph{Roadmap.}
The rest of the paper is organized as follows. Section~\ref{sec:related}
reviews related work about botnet, and security problems associated with the
emerging IoT device. Then, in Section~\ref{sec:mirai}, we present the
architecture of the Mirai botnet, as well as a typical vulnerable device
involved in the Mirai attacks. Section~\ref{sec:assum} summaries key features
of the Mirai botnet. The design and implementation are the proposed solution
are demonstrated in Section~\ref{sec:design} and Section~\ref{sec:imp}
respectively. Then we evaluate the effectiveness of the system in
Section~\ref{sec:eva}, followed by some discussions about the limitation the
work. Finally, Section~\ref{sec:con} concludes the paper.

\section{Related Work}
\label{sec:related}

\subsection{IoT Security}




The Internet of Things are emerging nowadays.
However, under the pressure of the time-to-market, manufacturers of these IoT devices pay little attention to the security and privacy enhancement for their products.
This kind of neglect results in serious hazards and attacks~\cite{iothazards22,iothazards11,Choccs2016,Sivaramanwisec2016,Yuhandling2015hotnets,uncovering2016das}.
To tackle these problems, more and more security researchers in academia and industry focus on IoT security.

 Pa et al.\cite{yiniotpot2015} analyzes the increasing threats against IoT devices and show that Telnet-based attacks that target IoT device have increased since 2014. It proposes IoT honeypot and sandbox on different CPU architectures to analyze the malware samples targeting Telnet-enabled IoT devices. 
Fernandes et al.\cite{smartthings16} provides the first depth of security analysis on the smart home programming platform, i.e. Samsung-owned SmartThings, and discovers the over-privilege problem.
This problem can make a malicious battery monitor SmartApp subscribe door lock PIN change event.
To solve this problem, Jia et al.\cite{contexiot17} propose a context-based permission system for appified IoT platforms that helps users perform effective access control.
Costin et al.\cite{Andrei2014firmware} performs a large-scale analysis of firmware images and discovered 38 previously unknown vulnerabilities in over 693 firmware images.
For example, some devices' firmware images hardcodes credentials.
Fernandes et al.\cite{flowfence2016earlence} present FlowFence, using information flow approach within application structure to prevent application misusing sensitive data in IoT devices.
In our work, we target to defeat Mirai in a simple, effective and manageable way.
It helps manufacturers build an implanter server to implant the virus expeller into vulnerable devices to immune Mirai's infection and eventually stop Mirai botnet's attack.

\subsection{Botnet}

Botnet has been the research topic since 2000s.
It can bring significant damages to the security of both individuals and businesses.
To counter botnet, there are usually two effective ways comprised of preventing botnoet's spreading and prevent botnet's attack.
Prevent botnet's spreading often needs the victim devices' users or administrators to fix their devices' vulnerabilities.
Without the capability of exploiting a specific vulnerability, the botnet cannot increase its size.
This results in the reduction of this botnet's attack power.

There are tow most common strategies to prevent botnet's attack.
They are disabling C\&C servers and sinkholing the attack traffic~\cite{botnetdefense1,Karami2016stress,Mirkovic2004taxonomy}.
Taking down C\&C servers can defeat the botnet, but just in a while.
Because botnet attacker can change C\&C servers periodically.
Besides, if the server are not in the scope of local law enforcement, it is not easy to do this.
This way also has a defect that even if C\&C servers are taken down, the botnet's zombies are still active in the wild which can be reused later by others. 
Sinkholing the attack traffic happens when one botnet is attacking a specific victim.
It requires the cooperation of upstream Internet service providers, such as ISP.
It also needs to identify the unique signature of this traffic to be filtered.
Our work defeat Mirai by preventing its spreading by implanting the virus implanter into the device.
In this way, Mirai cannot infect this specific device and its size cannot increase.

\section{Explaining Mirai}
\label{sec:mirai}

This section details Mirai's design, which is helpful in understanding the
proposed defense mechanism. To begin with, we introduce a Mirai-vulnerable
device.


\subsection{Mirai-vulnerable Device}
\label{sec:devices}

IoT devices subject to Mirai attacks are mainly IP Cameras and Digital Video
Recorders (DVR). These emerging devices are usually fragile and vulnerable,
because very few security measure is built into the system. To make things
worse, many of these IoT devices run naked in the Internet -- back-door
accounts can be used to remotely access the device with root privilege. In
this section, we dissect a Dahua digital video recorder of model DH-3004,
whose vulnerability statement is given at the end.

\paragraph{Hardware.}
This device has four BNC connector for video input, and two for audio input.
It also has two USB ports for connecting peripheral like flash disks and a
mouse. An Ethernet port is used to connect to the Internet and a VGA port is
used to connect a monitor. The device is powered by an ARM ARM926EJ-S
processor, with 43 MB of RAM. In addition, a hard disk can be plugged in a
SATA interface to record captured signals.

\paragraph{Software.}
The device runs \textit{HiLinux}, a tailored Linux distribution, integrated
with BusyBox as its toolset. The kernel is a legacy one of version 2.6.24.
By default, six TCP ports are open. They are  23, 80, 101, 102, 554, and 6623.
Some of them are used to telnet into the system for diagnose, while others are
used to connect a mobile app for remote control of the device. The system has
a root account with no password. Since the flash of the system is read-only,
there is no way to set up a new root password.

\paragraph{Vulnerability Statement.}
DH-3004 device has at least two ways for remote access, including telnet and
HTTP. Any one who can reach DH-3004 can login into the device with root
privilege through telnet access. As there is no way to set a root password,
and firmware cannot be updated, the device is highly vulnerable to Mirai
attacks.

\begin{figure}[t!]
\centering
\includegraphics{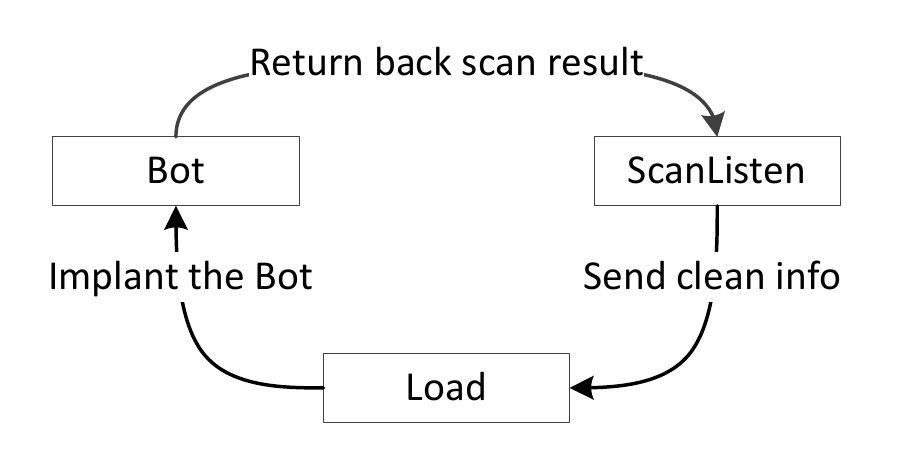}
\caption{Mirai's infection process}
\label{fig:mirai}
\end{figure}

\subsection{Mirai Analyses}

As a botnet malware, Mirai uses the client-server (CS) model to connect the
bot with a command and control (C\&C) server. However, compared with
traditional botnet malwares, Mirai is special in the infection phase. Instead
of infecting directly through the bots, Mirai botnet is only used to scan and
collect vulnerable devices and the server launches the actual attack to
implant botnets.


There are three modules in Mirai, a \texttt{Bot} module, a \texttt{ScanListen}
module, and a \texttt{Load} module. Figure \ref{fig:mirai} illustrates Mirai's
architecture, and the information flow and the relationship among these
modules. The Bot is a program running in victim devices. It scans other
devices on the Intent having the same  vulnerability. If it finds one, the
vulnerable device's information will be uploaded to the ScanListen module,
which runs in a pre-known server. The information includes login credentials,
device IP address and vulnerable ports, etc. After receiving the these
information, the ScanListen module sends it to the Load module, which runs in
the same server. The Load module then use these information to  infect the
target device and implant Bot.

\paragraph{The Bot Module.}
This is the attack module which perform DDoS the actual attacks to the victim
server. It is also used to scan the Internet to collect other vulnerable
devices, and collect their information. The Bot module implements features and
functions listed below.

\begin{enumerate}

\item \textbf{Preventing the device from rebooting}. Mirai's bot only exits in
the memory. If the device reboots, bot disappears. Therefore, in order to
prevent this from happening, bot writes the request command ``0x80045704'' to
the ``watchdog'' of the device to disable rebooting in face of a system hang
up.

\item \textbf{Process hiding.} Mirai uses a random string to hide its process
name.

\item \textbf{Preventing second infection.}
The Bot module opens the port 48101 and binds to this port. If another bot
wants to bind this port, it would detect this event. In this way, mirai
ensures there is only one bot running on the target device.

\item \textbf{Blocking ports.} Mirai closes port 23(telnet), 22(ssh), 80(http)
to block other botnet malware's attack.

\item \textbf{Experlling other malwares.}
Mirai's bot module scans the system to find the fingerprint of other malwares.
With root previlege, Mirao is able kill other malware processes.

\item \textbf{Device Scan.} Mirai bot periodically, scans neighboring devices
to discover vulnerable ones. However, it excludes those device whose IP
addresses belong to the General Electric Company, Hewlett-Packard Company, US
Postal Service, etc. as depicted in listing \ref{lst:ipfilter}. If a
vulnerable device is fount, the bot module sends back this device's
information to the ScanListen module. Bot uses brute-force to scan other
devices. Namely, Bot hard-codes 62 pairs of back-doored user name and password
pair. Listing \ref{lst:userpassword} lists the revealed username-password
list. These string is further obfuscated by a simple xor operation by
``0xDEADBEEF''.


\item \textbf{DDoS Attack.} The bots connect to the C\&C server and wait for
the command to attack the target server.

\end{enumerate}

\begin{lstlisting}[caption={IP filter in bot}, label={lst:ipfilter}, style=c]
...
while (o1 == 127 ||                                       // 127.0.0.0/8      - Loopback
      (o1 == 0)  ||                                       // 0.0.0.0/8        - Invalid address space
      (o1 == 3)  ||                                       // 3.0.0.0/8        - General Electric Company
      (o1 == 15  || o1 == 16) ||                          // 15.0.0.0/7       - Hewlett-Packard Company
      (o1 == 56) ||                                       // 56.0.0.0/8       - US Postal Service
      (o1 == 10) ||                                       // 10.0.0.0/8       - Internal network
      (o1 == 192 && o2 == 168) ||                         // 192.168.0.0/16   - Internal network
      (o1 == 172 && o2 >= 16 && o2 < 32)  ||              // 172.16.0.0/14    - Internal network
      (o1 == 100 && o2 >= 64 && o2 < 127) ||              // 100.64.0.0/10    - IANA NAT reserved
      (o1 == 169 && o2 > 254) ||                          // 169.254.0.0/16   - IANA NAT reserved
      (o1 == 198 && o2 >= 18 && o2 < 20)  ||              // 198.18.0.0/15    - IANA Special use
      (o1 >= 224) ||                                      // 224.*.*.*+       - Multicast
      (o1 == 6 || o1 == 7 || o1 == 11 || o1 == 21 || o1 == 22 || o1 == 26 || o1 == 28 || o1 == 29 || o1 == 30 || o1 == 33 || o1 == 55 || o1 == 214 || o1 == 215)  // Department of Defense
);
...
\end{lstlisting}

\begin{lstlisting}[caption={Hardcoded usernames and passwords in bot}, label={lst:userpassword}, style=c]
// Set up passwords
add_auth_entry("\x50\x4D\x4D\x56", "\x5A\x41\x11\x17\x13\x13", 10);      // root  xc3511
add_auth_entry("\x50\x4D\x4D\x56", "\x54\x4B\x58\x5A\x54", 9);           // root  vizxv
add_auth_entry("\x50\x4D\x4D\x56", "\x43\x46\x4F\x4B\x4C", 8);           // root  admin
add_auth_entry("\x43\x46\x4F\x4B\x4C", "\x43\x46\x4F\x4B\x4C", 7);       // admin admin
add_auth_entry("\x50\x4D\x4D\x56", "\x5A\x4F\x4A\x46\x4B\x52\x41", 5);   // root  xmhdipc
add_auth_entry("\x50\x4D\x4D\x56", "\x46\x47\x44\x43\x57\x4E\x56", 5);   // root  default
add_auth_entry("\x50\x4D\x4D\x56", "", 4);                               // root  (none)
...
\end{lstlisting}

\paragraph{The ScanListen Module.}
This module is responsible for sending the information collected by the bots
to the Load module. The format is ``IP-Address:Port'' and
``User-Name:Password''.

\paragraph{The Load Module.}
This module gets the input from ScanListen and performs attack against each
vulnerable devices. The critical steps are as following.

\begin{enumerate}

\item Login to the target device through exposed vulnerabilities.

\item Make sure the target device has BusyBox installed.

\item Find a read-writable directory.

\item Copy ``/bin/echo'' to the read-writable directory.

\item Implant bot through ``echo'', ``wget'' or ``tftp'' commands by connecting a server.

\item Execute the bot program in memory.

\item Delete the bot program in the file system.

\end{enumerate}


\section{Dissection of Mirai}
\label{sec:assum}


This section dissects Mirai, lists some key features in the spread of Mirai,
and argues some assumptions regarding the capabilities of customers.

Mirai is widely deployed in the wild~\cite{mirai-deploy} and there are thousands of Mirai
bots on the Internet scanning other vulnerable devices. The devices infected
by Mirai have no way to update firmware. In addition, most of them operate on
a temporary in-memory file system, which loses its contents after power off.
By keeping track of each sold device, the device manufacturer can contact its
customers, e.g., by the email. The customer may not a power user, who has
advanced computer skills. We assume the customer is only capable of following
the product manual to configure the device and power on/off the device.

Although Mirai has an effective way to scan devices, the IP addresses are
random chosen. There should be a time window when the device is in clean state
after rebooting.

The way Mirai deploys the botnet is straightforward. It first remotely
exploits the weak-password vulnerability, then implants the bot and waits for
the command to attack the target server. In this process, once implanted into
a device, the bot closes several ports to disable remote access to the device
originated other malware. The bot also deletes the malicious executable binary
on the file system to conceal its presence. 

From the aforementioned analyses on Mirai botnet, we figure out two key
factors in the infection of Mirai. They are the remote access and the
hard-coded password vulnerability. Therefore, in order to defeat Mirai, there
are two effective ways, including fixing the password vulnerability and
closing the ports for remote access. As the device's firmware cannot be
updated, the password vulnerability cannot be fixed permanently. In addition,
the device may not support changing password even temporarily. On the other
hand, although closing the ports for remote access cannot persist either,
through collaboration between the customer and manufacture, it is possible to
minimum the time of vulnerable devices with open ports exposed on the Internet.

\section{Desgin}
\label{sec:design}


The key idea of our solution is to use a ``tit for tat'' strategy to defeat
Mirai. That is, the manufacturer deploys a Mirai-like botnet system to implant
virus expeller into the vulnerable devices and close the  ports for remote
access to defeat Mirai automatically. Before illustrating the system, we first
discuss the challenges which can help understand the system design.

\subsection{Challenges}
\label{sec:challenge}
The basic idea of our solution is also implanting a ``white'' Mirai to the
vulnerable device. The first challenge is how to control the infection range
and the influence on the normal operations of the devices. From the analysis
above, Mirai uses bots to scan other devices and implants bots into the
vulnerable devices through the Load module automatically. When deploying the
Mirai-like system, we should consider the overhead that can bring to the
Internet and the devices. For example, these IP camera and DVR devices are
usually resource-constrained. Exhausting device's resource to defeat Mirai  is
not acceptable for the customers. Besides, if a manufacturer scans the
Internet to discover the vulnerable devices of its own, this kind of spam
stream may block some network streams. Therefore, how to control the
Mirai-like system to make it effective and bring minimum influence on the
Internet is very important.

The second challenge is how to ensure our ``white'' Mirai to be implanted into
the target device effectively. In section~\ref{sec:assum}, we assume Mirai has
been widely deployed in the wild and there are thousands of bots to find the
vulnerable devices. In fact, the botnet deployed by the attacker is very
large~\cite{mirai-deploy}. Mirai uses bot to scan the devices, so the larger the botnet
is, the faster Mirai can infect a new device. If the implanting of the
manufacturer  is not as efficient as the wild Mirai, Mirai bot can infect a
device earlier than ours. In this case, as the ports has been closed by
Mirai's bot, our ``white'' Mirai cannot be implanted at all.

\begin{figure}[t!]
\centering
\includegraphics[scale=0.65]{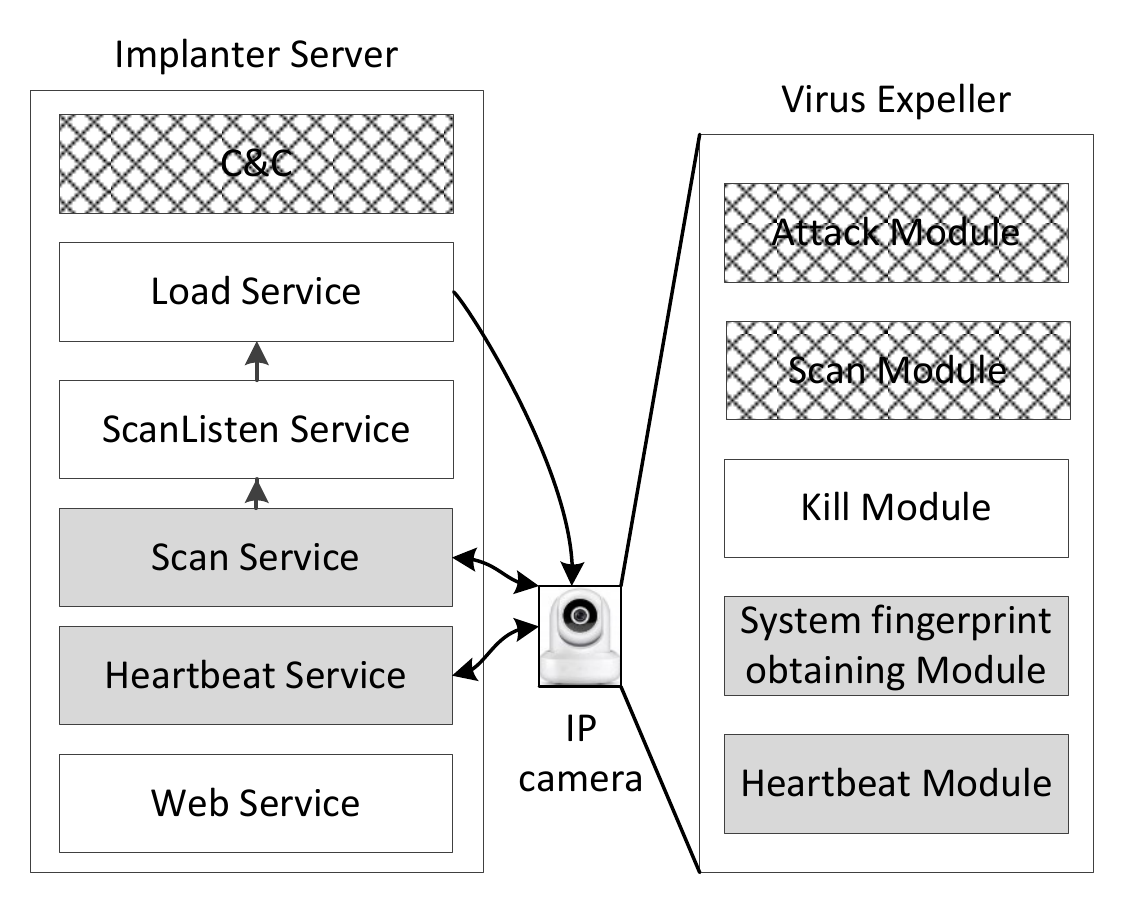}
\caption{The Mirai-like system to defeat Mirai}
\label{fig:design}
\end{figure}

\subsection{System Architecture}

Figure \ref{fig:design} depicts the system design of our ``white'' Mirai and
its difference with the original Mirai. The gridded modules are those removed
from the original Mirai. As our system derives from Mirai, we must remove the
code and modules designed to perform DDoS attacks to prevent our benign system
is abused. The gray modules are these newly added into the system. The two
most important modules include a virus expeller and a implanter server, which
run on the vulnerable device and the manufacture maintained server
respectively.


\paragraph{The Virus Expeller.} Once implanted into a vulnerable device, the
virus expeller expels the infection of Mirai's bot. The virus expeller
inherits code from Mirai's bot, but distinguishes itself from Mirai bot by
removing the attack module and scan module. Besides, as depicted in
figure~\ref{fig:design}, the virus expeller also adds several functions, such
as fingerprint obtaining module and heart-beating module. Fingerprint
obtaining module is responsible for collecting device's information, such as
its unique ID. This module is used to avoid the legal problem. In particular,
if the virus expeller is injected into a vulnerable device which is produced
by another manufacture, without acknowledgment from the customer, legal
problems arise. Therefore, the fingerprint obtaining module can make sure the
infected device belongs to the exact manufacturer.

The heart-beating module is a client program for the heartbeat service and reports
aliveness to it in the implanter server periodically. More details about
this heartbeat functionality will be described in section~\ref{sec:implanter}.

The virus expeller's core part is a blocking module. It closes the ports for
remote access as soon as it is implanted into a device. However, there is a
time window between this virus expeller is executed and the ports are closed.
This time window allows original Mirai to infect this specific device. Thus,
the blocking module kills Mirai's bot process if one is found.

As Mirai-like system deploys virus expellers in the resource-constraint
devices, the resource used in these devices should be restricted. Otherwise,
the user may complain about this heavy-weighted protection method. Because the
scan module in the original Mirai's bot consumes a lot of resources, in order
to limit the influence on the device, instead of performing scan on the
device, we move this function to the implanter server. Another benefit of
removing the scan function is to avoid generating  much spam stream which has
a bad influence on the Internet.

\paragraph{The Implanter Server.}
\label{sec:implanter}
The implanter server runs five  programs, including Load module, ScanListen
module, Heartbeat module, Scan module, and HTTP module.

The scan module inherits code from Mirai's scan module. It is used to scan the
Internet to find the vulnerable devices. However, in order to implant the
virus expeller only to the products of a specific manufacturer, the scan
module keeps this specific manufacturer's information. Furthermore, this
module is extended and can scan an IP address range or a specific IP address
to find vulnerable devices.

The load module and the scanlisten module have no difference with Mirai's load
and scanlisten modules. They are responsible for collecting the vulnerable
devices' information from the scan service and implanting the virus expeller
into these devices. The load module is also used for downloading the
executable payloads from a HTTP module on the server.

The heartbeat module monitors each virus expeller's aliveness and re-implant
them if some accident happens, such as devices' unplanned power-on and
power-off, as this would clear the virus expeller. There may be a concern in
heartbeat service's usage, i.e. whether these virus expellers may results in a
DDoS attack for this implanter server. But according to the material exposed
by Mirai's authors~\cite{mirai-post}, 2\% CPU can support 400k bots. Therefore, this is
not a problem.


\begin{figure}[t!]
\centering
\includegraphics[scale=0.65]{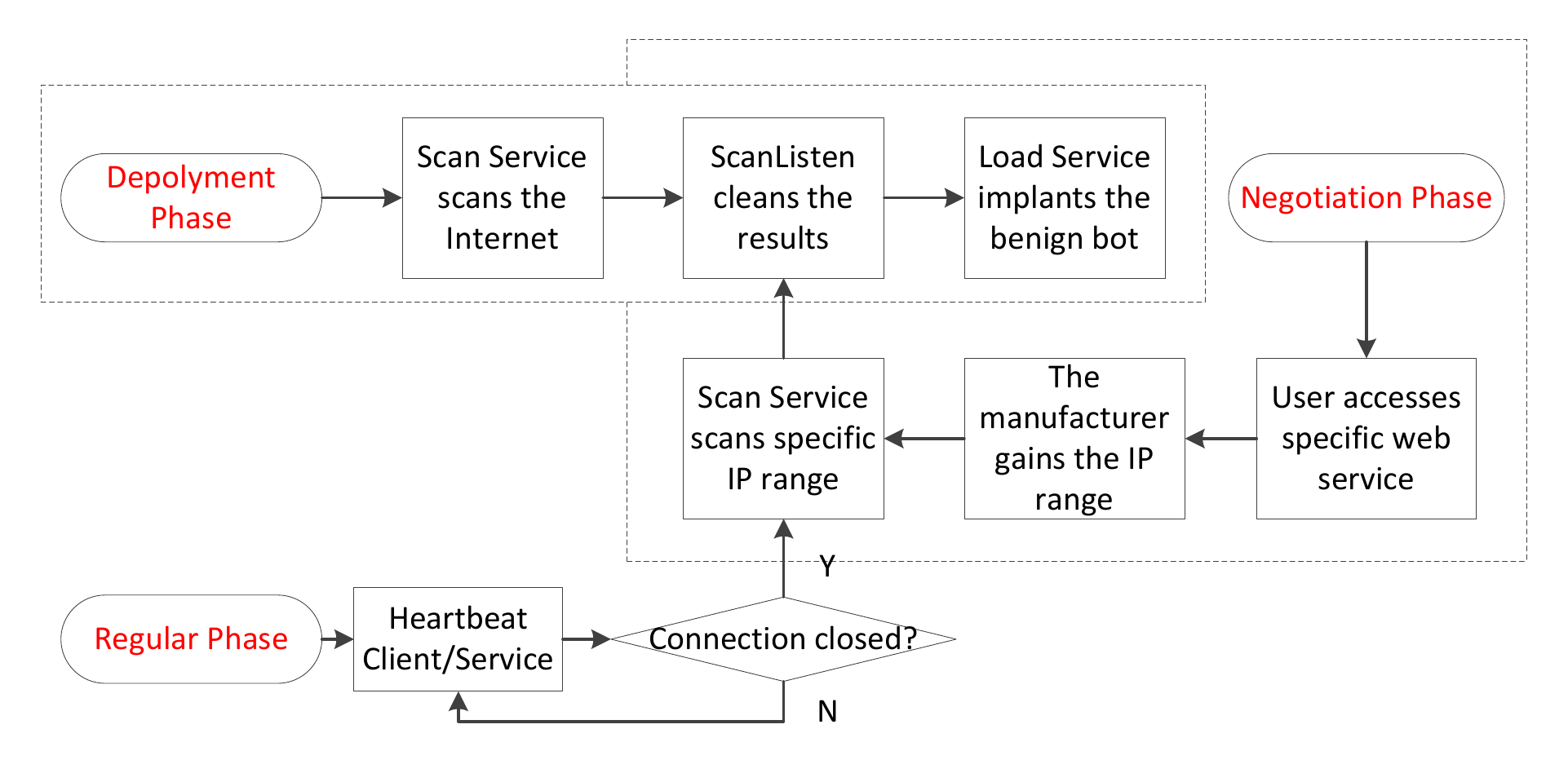}
\caption{Three Phases in the Proposed Mirai-like System}
\label{fig:phases}
\end{figure}

\subsection{Work Flow}

The operation of our system has three phases, deployment phase, regular phase
and negotiation phase. Figure \ref{fig:phases} illustrates how three phases
work together to defeat Mirai attacks.

\paragraph{Deployment Phase.} In this phase, the manufacturer deploys the
system by firstly running the scan service to find vulnerable devices and
implant the virus expeller into them. This phase is a basic phase which is
simple and straightforward. Its effectiveness depends on how quickly the scan service can
find vulnerable devices. In the real world, this phase may be optimistic.
Because in our assumption, Mirai is already deployed in the wild and all the
devices on the Internet may be infected by it. However, there should be some
devices which are powered on and the time window is long enough for this scan
service to find them and implant the virus expeller.

\paragraph{Regular Phase.} The manufacturer is not involved in this phase. In
this phase, the heartbeat client in the virus expeller sends back a heartbeat
to the heartbeat service in the implanter server periodically. At the same
time, the heart-beating service has a timer for each virus expeller and
continuously checks them. If the timer exceeds a specific value, it means the
related virus expeller loses connection with the server. In this way, the
manufacturer can know whether a virus expeller has successfully ran on a
specific device. If this device is accidentally rebooted, the manufacturer can
know exactly which device's virus expeller loses connection with the heartbeat
service. Then the heartbeat service can invoke the scan service to scan a
specific IP address range deduced from the specific device information
obtained previously, and can re-implant the virus expeller into the device
afterwards.

\paragraph{Negotiation Phase.} In the last two phases, we implicitly assume
that our Mirai-like system can implant the virus expeller into a target
device. However, in practice, as discussed in section~\ref{sec:challenge},
Mirai botnet has a large number of bots and can already infect a target
device. If Mirai bot has already been in a device, no virus expeller can be
implanted into this device. Because Mirai bot firstly closes the ports for
remote access, there is no other way to access this device. In order to handle
this situation, the customer must collaborate with the manufacturer to clear
Mirai and implant the virus expeller. In this phase, the customer and the
manufacturer firstly agree on a specific time. At that time slot, the user
uses a computer, to access a special web page operated by manufacturer, and
then reboots the vulnerable device. The manufacturer can get a clue about the
IP range this specific user's device is in. As a result, the manufacturer can
leverage the scan service to quickly scan the IP range to locate this
vulnerable device and implant the virus expeller before Mirai can infect this
device. As the Mirai botnet does not know the time negotiated by the customer
and the manufacture, our solution can effectively block spread of original
Mirai. Furthermore, according to the experiment done by~\cite{mirai-infection}, it costs
around 98 seconds for Mirai  botnet to infect a device after being connected
to the Internet. So there is little chance for Mirai to infect this device at
the very moment that the customer reboots the device.

With regard to the IP range, we may need to consider different circumstances
in reality. If a device is in the private network and does not have a public
IP address, it seems the Mirai-like system cannot infect this device directly.
However, the same applies to Mirai. Moreover, some devices leverage UPnP
(Universal Plug and Play) protocol to do NAT traversal and expose themselves
on the Internet for user convenience. As a result, the Mirai botnet can
directly connect to this device through the public IP address.

\section{Implementation}
\label{sec:imp}


Our implementation is based on the open-sourced Mirai code~\cite{mirai-code}. Mirai's
code base is more than 9K Lines Of Code (LOC), including C/C++, Go and Bourne
Shell. As we remove attack modules from Mirai and add heartbeat client/server
modules, etc., the code base of our implementation is about 6K LOC. In this
section, we only describe the functions newly added into the virus expeller
and the implanter server.

\subsection{The Virus Expeller}

Two functions are added into the virus expeller, They are the fingerprint
obtaining module and the heart-beating module. The former gets system
information from pseudo files, such as ``/proc/cpuinfo'', ``/proc/hiversion'',
etc. These information is used to determine whether this device is  the
product of a specific manufacturer.

The heart-beating module is simple. It sends back a magic number
``0xE84Eb1C8'' to the heartbeat service every minute to report the virus
expeller's aliveness.

\subsection{The Implanter Server}

The implanter server has been added two more services, including a scan service
and heartbeat service. The scan service inherits code from Mirai's bot scan
module. It is modified to scan a specific manufacturer's products with
specific username-password pairs. This can accelerate the scan speed and to
some extent exclude other manufacturer's products. Furthermore, in the
original Mirai's bot, the scan module only get random IP addresses to scan. In
ours, we only need to scan a specific IP range obtained from the load module.

The heartbeat module is the server process for the heartbeat client in the
virus expeller. It records each virus expeller's IP address, refreshes the
timer for each virus expeller, and checks the timer periodically. In our
design, the heart-beating module checks the timer to see whether it exceeds 70
seconds. If a timer exceeds the time limit, this module will invoke the scan
service to scan that specific IP address and the related IP address range to
relocate this device, in case it is accidentally rebooted.


\section{Evaluation}
\label{sec:eva}


We evaluate the functionality of the proposed Mirai-like system using the
device mentioned in section~\ref{sec:devices}. The implanter server runs
in a Debian 8 Linux operating system which is powered by an Intel i7-4790
processors with 2GB memory.

We are interested in three functions.
The first is whether the proposed system can implant the virus expeller into a target device through this process: Scan $\to$ ScanListen $\to$ Load $\to$ Web $\to$ Virus Expeller (in the target device) $\to$ Heartbeat.
If the virus expeller is implanted into a device, the heartbeat service can receive the messages sent by it.
Using wireshark running on a PC which monitors the network packages,
we measured the time taken to implant the virus expeller since the scan service starts to scan this device.
On average, it takes about ~10 seconds, which is much faster than the wild Mirai does.
After that, the heartbeat can periodically receive a message sent from the virus expeller which shows that this system works.

The second is whether the virus expeller can defeat Mirai's infection once
being implanted into a device. We run this experiment by implanting the virus
expeller into this target device firstly and then run the scan service to scan
this specific device to find whether the load service can receive any
information. As our scan service, scanlisten service and load service inherit
code from Mirai, if our Mirai-like system can scan the device and return
related information to the scanlisten service, this means the target device
can be infected by Mirai even there is a virus expeller inside. From load
service's log message, we can directly find the virus expeller is implanted
into this device and no more message is generated after that. Besides, we try
to use telnet to connect to this device and fail to connect into this device.
Therefore, once being implanted into a device, the virus expeller can succeed
in defeating Mirai by closing related remote access ports.

The third is whether the implanter server can re-implant the virus expeller
into this specific device, if the device is accidentally rebooted. We run this
experiment by rebooting the device and monitor the log of the heartbeat
service. From the log, we can see a list of operations. First, the virus
expeller stops sending back the aliveness message. Second, the heartbeat
service invokes scan service to scan the target IP address range. Third, the
scan service finds the target device and the virus expeller is implanted by
the load service.

\section{Discussion}
\label{sec:dis}



\subsection{Limitation of the Present Work} The proposed work implants a virus
expeller into a target device to defeat the Mirai botnet malware. Our system
includes three phases, i.e. deployment phase, regular phase and negotiation phase. The
deployment phase is a basic phase which leverage the scan service to find vulnerable
devices in the Internet and the load service to implant the virus expeller. The regular phase is
based on the heartbeat module and scans service to keep the virus expeller
implanted in the device. These two phases potentially assume the Mirai-like
system can implant the virus expeller into the target device. But in the wild,
there may be few devices left intact which result in the efficiency of these
two phases. The negotiation phase involves the device's user, which introduces determinacy into the system.
However, the user may not be cooperative, which limits the
efficiency of our system.

The proposed work only defeats the original Mirai botnet malware for now. This
malware' variants may use different ports or use different vulnerabilities to
infect the target devices. For different ports, this system can add them
into the virus expeller's port list which is easy to deploy. For  new
vulnerabilities, our system can also defend the infection by closing the
specific ports used in these vulnerabilities. However, a port closed by the
virus expeller may be a functional port used by the device. When it is
closed, the device's function is influenced. For example, port 80 is used by
the DH-3004 device to show the web interface. If it is closed, the user can no
longer access it. 


With regards to the potential threat to the different participants, such as the
user, the manufacturer, the ISP, etc., the only concern is the influence on
the Internet and the device. As the attack module has been remove from Mirai,
our system cannot be abused. Besides, this system has removed the scan
module from the virus expeller. However, despite of the above mitigation, there
may still exist some influences on the Internet and the device.


There are also legal concerns related with the present work~\cite{mirai-law}. Although our system limit
the infection devices within this specific manufacturer's products, the way it
uses is still not decent. Furthermore, different countries may have different
laws to justify this behavior. Thus, our system only works if the appropriate
legal framework is in place to allow this behavior.

\subsection{Future Work} The present work solves the problem that the Mirai botnet
malware exploits the weak password vulnerability in IoT devices to deploy
botnet and perform DDoS attacks. It leverages Mirai's code base and implants the
virus expeller into a target device to defeat Mirai. At present, our system
only works only if the target device is clean, i.e. Mirai has been cleared by
rebooting or this device has not been infected by Mirai at that time. In the future,
we will assume that the device has already been infected by Mirai.
One potential solution is to discover Mirai's vulnerability and exploit
this vulnerability to disarm Mirai.


\section{Conclusion}

\label{sec:con}
We have presented a new mechanism to aid manufacture in amending the
Mirai-vulnerable devices in the market. Different from recalling or customer
services, in which customers are unwilling to cooperate, our solution needs
minimal involvement of customers. Different from the ``white'' Mirai proposed
in~\cite{anti-mirai}, our solution does not need brute-force scan to infect others, and
can deterministically patch the compromised device.

A proof-of-concept prototype has been implemented. Experimental results show that the
proposed solution is both simple and effective. Given the great number of
Mirai-vulnerable devices in the wild, our solution provides an attractive path
towards mitigating the threats from Mirai, until all the vulnerable devices
are retired. Since our solution requires close cooperation with the
manufactures, we plan to contact the involved manufactures to further carry
out our solution in real world.

\bibliographystyle{abbrv}
\bibliography{bib/myrefs}

\begin{thebibliography}{10}

\bibitem{mirai-recall}
Chinese firm recalls camera products linked to massive ddos attack.
\newblock
  http://www.pcworld.com/article/3133962/chinese-firm-recalls-camera-products-linked-to-massive-ddos-attack.html.

\bibitem{mirai-ddos}
Ddos attack that disrupted internet was largest of its kind in history, experts
  say.
\newblock
  https://www.theguardian.com/technology/2016/oct/26/ddos-attack-dyn-mirai-botnet.

\bibitem{mirai-law}
Federal computer intrusion laws.
\newblock
  http://iwar.org.uk/law/resources/doj/federal-computer-intrusion-laws.htm.

\bibitem{mirai-post}
Form post about mirai.
\newblock
  https://github.com/jgamblin/Mirai-Source-Code/blob/master/ForumPost.md.

\bibitem{mirai-first}
Hacked cameras, dvrs powered today¡¯s massive internet outage.
\newblock
  https://krebsonsecurity.com/2016/10/hacked-cameras-dvrs-powered-todays-massive-internet-outage/.

\bibitem{mirai-deploy}
The hackers popopret and bestbuy are offering a ddos-for-hire service
  leveraging a mirai botnet composed of around 400,000 compromised devices.
\newblock
  http://securityaffairs.co/wordpress/53824/cyber-crime/mirai-botnet-ddos-for-hire.html.

\bibitem{iothazards22}
Hacking the human heart.
\newblock http://bigthink:com/future-crimes/hackingthe-human-heart.

\bibitem{mirai-infection}
Mirai infection experiment.
\newblock https://twitter.com/ErrataRob/status/799556482719162368.

\bibitem{mirai-code}
Mirai source code.
\newblock https://github.com/jgamblin/Mirai-Source-Code.

\bibitem{botnetdefense1}
Simda: A botnet takedown.
\newblock
  http://blog.trendmicro.com/trendlabs-security-intelligence/simda-a-botnet-takedown/.

\bibitem{iothazards11}
Team of hackers take remote control of tesla model s from 12 miles away.
\newblock
  https://www:theguardian:com/technology/2016/sep/20/tesla-model-s-chinese-hack-remote-control-brakes.

\bibitem{Choccs2016}
K.-T. Cho and K.~G. Shin.
\newblock Error handling of in-vehicle networks makes them vulnerable.
\newblock In {\em Proceedings of the 2016 ACM SIGSAC Conference on Computer and
  Communications Security}, CCS '16, pages 1044--1055. ACM, 2016.

\bibitem{Andrei2014firmware}
A.~Costin, J.~Zaddach, A.~Francillon, and D.~Balzarotti.
\newblock A large-scale analysis of the security of embedded firmwares.
\newblock In {\em 23rd USENIX Security Symposium (USENIX Security 14)}, pages
  95--110. USENIX Association, 2014.

\bibitem{uncovering2016das}
A.~K. Das, P.~H. Pathak, C.-N. Chuah, and P.~Mohapatra.
\newblock Uncovering privacy leakage in ble network traffic of wearable fitness
  trackers.
\newblock In {\em Proceedings of the 17th International Workshop on Mobile
  Computing Systems and Applications}, HotMobile '16, pages 99--104. ACM, 2016.

\bibitem{smartthings16}
E.~Fernandes, J.~Jung, and A.~Prakash.
\newblock {S}ecurity {A}nalysis of {E}merging {S}mart {H}ome {A}pplications.
\newblock In {\em Proceedings of the 37th {IEEE} Symposium on Security and
  Privacy}, May 2016.

\bibitem{flowfence2016earlence}
E.~Fernandes, J.~Paupore, A.~Rahmati, D.~Simionato, M.~Conti, and A.~Prakash.
\newblock Flowfence: Practical data protection for emerging iot application
  frameworks.
\newblock In {\em 25th USENIX Security Symposium (USENIX Security 16)}, pages
  531--548. USENIX Association, 2016.

\bibitem{anti-mirai}
J.~A. Jerkins.
\newblock Motivating a market or regulatory solution to iot insecurity with the
  mirai botnet code.
\newblock In {\em 2017 IEEE 7th Annual Computing and Communication Workshop and
  Conference (CCWC)}, pages 1--5, Jan 2017.

\bibitem{contexiot17}
Y.~J. Jia, Q.~A. Chen, S.~Wang, A.~Rahmati, E.~Fernandes, Z.~M. Mao, and
  A.~Prakash.
\newblock {ContexIoT: Towards Providing Contextual Integrity to Appified IoT
  Platforms}.
\newblock In {\em 21st Network and Distributed Security Symposium}, 2017.

\bibitem{Karami2016stress}
M.~Karami, Y.~Park, and D.~McCoy.
\newblock Stress testing the booters: Understanding and undermining the
  business of ddos services.
\newblock In {\em Proceedings of the 25th International Conference on World
  Wide Web}, WWW '16, pages 1033--1043. International World Wide Web
  Conferences Steering Committee, 2016.

\bibitem{Mirkovic2004taxonomy}
J.~Mirkovic and P.~Reiher.
\newblock A taxonomy of ddos attack and ddos defense mechanisms.
\newblock {\em SIGCOMM Comput. Commun. Rev.}, 34(2):39--53.

\bibitem{yiniotpot2015}
Y.~M.~P. Pa, S.~Suzuki, K.~Yoshioka, T.~Matsumoto, T.~Kasama, and C.~Rossow.
\newblock Iotpot: Analysing the rise of iot compromises.
\newblock In {\em 9th USENIX Workshop on Offensive Technologies (WOOT 15)}.
  USENIX Association, 2015.

\bibitem{Sivaramanwisec2016}
V.~Sivaraman, D.~Chan, D.~Earl, and R.~Boreli.
\newblock Smart-phones attacking smart-homes.
\newblock In {\em Proceedings of the 9th ACM Conference on Security \&\#38;
  Privacy in Wireless and Mobile Networks}, WiSec '16, pages 195--200. ACM,
  2016.

\bibitem{Yuhandling2015hotnets}
T.~Yu, V.~Sekar, S.~Seshan, Y.~Agarwal, and C.~Xu.
\newblock Handling a trillion (unfixable) flaws on a billion devices:
  Rethinking network security for the internet-of-things.
\newblock In {\em Proceedings of the 14th ACM Workshop on Hot Topics in
  Networks}, HotNets-XIV, pages 5:1--5:7. ACM, 2015.

\end{thebibliography}

\end{document}